\documentclass[twocolumn,amssymb,fleqn]{revtex4} 
\usepackage{epsfig,amssymb,amsmath,graphicx,subfigure,hyperref  }  
\usepackage{color, setspace} 
\usepackage{multirow}
\usepackage{tabularx}

\newcommand{\be}{\begin{eqnarray}}   
\newcommand{\ee}{\end{eqnarray}}

\begin{document}

\title{Dressed Active Particles in Spherical Crystals}
\author{Zhenwei Yao} 
\affiliation{Department of Physics and Astronomy, and Institute of Natural
Sciences, Shanghai Jiao Tong University, Shanghai 200240 China}
\begin{abstract}
We investigate the dynamics of an active particle in two-dimensional spherical crystals, which
provide an ideal environment to illustrate the interplay of active particle and
crystallographic defects. A moving active particle is observed to be surrounded
by localized topological defects, becoming a dressed active particle. Such a
physical picture characterizes both the lattice distortion around the moving
particle and the healing of the distorted lattice in its trajectory. We
find that the dynamical behaviors of an active particle in both random and
ballistic motions uniformly conform to this featured scenario, whether the
particle is initially a defect or not. We further observe that the defect
pattern around a dressed ballistic active particle randomly oscillates
between two well-defined wing-like defect motifs regardless of its speed. The
established physical picture of dressed active particles in this work partially
deciphers the complexity of the intriguing nonequilibrium behaviors in active
crystals, and opens the promising possibility of introducing the activity to
engineer defects, which has strong connections with the design of materials.
\end{abstract}
\maketitle

\section{Introduction}

Ubiquitous nonequilibrium condensed matter systems exhibit a wealth of
intriguing properties not found in the equilibrium zone.~\cite{vicsek1995novel,
zwanzig2001nonequilibrium, marchetti2013hydrodynamics, ni2013pushing} From
highly coherent collective motions in moving animals,~\cite{parrish1997animal,
toner1998flocks}  bacterial suspensions,~\cite{zottl2016emergent,
zhang2010collective}  living cells,~\cite{rappel1999self,trepat2009physical,
wang2011spontaneous} and living liquid crystals~\cite{zhou2014living} to
emergent ordered structures developed in granular matters in
vibration,~\cite{aranson2006patterns, narayan2007long}
colloids,~\cite{zottl2016emergent, mognetti2013living}
nanoparticles,~\cite{paxton2004catalytic} soft particles,~\cite{henkes2011active} and active
spinners,~\cite{nguyen2014emergent, spellings2015shape}  these seemingly
distinct systems have a unifying characteristic that they are composed of
self-driven active units and have been known as active matters or living
matters.~\cite{vicsek1995novel, schweitzer2007brownian,
marchetti2013hydrodynamics} A large variety of these nonequilibrium properties
can be well rationalized in a unified physical model by endowing interacting
constituent particles with activity.~\cite{schweitzer2007brownian,
vicsek2012collective, marchetti2013hydrodynamics} Through various mechanisms
such as external electric or magnetic field,~\cite{tierno2008controlled}
light,~\cite{palacci2013living, palacci2013photoactivated,zhang2014accelerated}
mechanical vibration,~\cite{aranson2006patterns} chemical
reaction,~\cite{paxton2004catalytic, chaudhuri2011spatiotemporal} and
biological activity,~\cite{sanchez2012spontaneous} $\it etc.$, the energy input
through individual active particles drives the system out of equilibrium and
produces various ordered dynamic states and even biomimetic
behaviors.~\cite{marchetti2013hydrodynamics} Recent studies in both
polar~\cite{shi2013topological, giomi2013defect,sknepnek2015active} and
apolar~\cite{schaller2013topological} active systems, depending on whether the
constituent particles have a head and a tail, have shown the crucial role of
singular points known as topological defects~\cite{penrose1965dermatoglyphic}
in organizing active particles to move in a highly coherent fashion.
Remarkably, the combination of topological defects and activity can produce a
myriad of dynamical states as demonstrated in a recent experiment, where tunable
periodic oscillation of the defects in the active nematic vesicle has been
directly observed.~\cite{keber2014topology}  These studies suggest that the
complexity in the intriguing nonequilibrium behaviors arising from activity may be
characterized by the dynamics of a few topological defects. It is
therefore of interest to study the interplay of active particle and
topological defects to enhance our understanding of the activity induced
complex dynamics in active matters.

Spherical crystal is an ideal model system to study the physics of topological
defects.~\cite{bowick2002crystalline, bausch2003grain,bowick2006crystalline} In
a spherical crystal, particles are confined on the surface of sphere to
constitute a two-dimensional crystal lattice. Topological defects are
inevitable in two-dimensional crystalline order confined on spherical geometry
due to the topological constraint.~\cite{struik88a}  These defects provide the
unique opportunity to investigate the interplay of activity and topological
defects.  Disclinations are the elementary topological defects in
two-dimensional hexagonal lattices.~\cite{chaikin2000principles} An $n-$fold
disclination is a vertex whose coordination number $n \ne 6$. A topological
charge of $q=6-n$ can be assigned to an $n-$fold disclination. Note that one
should distinguish between topological charge and electric charge associated
with a particle.  All the particles in spherical crystal are electrically
charged, while a particle is topologically charged if its coordination number
is deviated from six.  According to the elasticity theory of topological
defects, disclinations of the same sign repel and unlike signs
attract.~\cite{bowick2006crystalline} Euler's theorem states that the total
topological charge in any spherical crystal is 12.~\cite{struik88a} It is
important to note that a particle in the spherical crystal can be assigned an
active force, becoming an active particle. A particle can also be a
disclination if its coordination number is deviated from 6. The double role of
a particle in the spherical crystal opens the possibility of moving a
disclination by assigning activity on it.

In this work, we introduce an active particle in spherical crystal, where
topological defects are inevitable. A spherical crystal can be experimentally
realized by confining electrically charged particles on sphere, which can
spontaneously form a hexagonal lattice with scattered disclinations under the
Coulomb potential.~\cite{bausch2003grain} We consider dynamics of the
particles in the overdamped regime described by the Langevin
equation.  The objective of this work is to understand the
activity induced nonequilibrium physics in terms of the elements of topological
defects. We prepare the initial state of the spherical crystal with the
simplest configuration of evenly distributed 12 isolated 5-fold disclinations.
We first numerically observe the intermediate hexatic phase in the noise driven
melting of the spherical crystal. It is comparable with the scenario of the
dislocation-mediated melting theory of two-dimensional crystals proposed by
Kosterlitz, Thouless, Halperin, Nelson and Young (KTHNY theory), suggesting the
reliability of our simulations.~\cite{kosterlitz1973ordering,
halperin1978theory, nelson1979dislocation, young1979melting,
strandburg1988two}

At a low level of noise below the melting point, we impose an active force on a
particle in the spherical crystal and track its dynamics and the resulting
adjacent lattice distortion. Such a lattice distortion is well represented by
the underlying topological defect structure via the triangulation of the
particles on sphere.  Extensive simulations show that an active particle always
carries disclinations around it in both random and ballistic motions, whether
the particle is initially a disclination or not. We name such a compound
structure of active particle and surrounding disclinaitions as a dressed active
particle.  For an initially disclinational active particle, we numerically
observe its splitting into an isolated disclination and a dressed active
particle with zero topological charge. The topological charge of the active
particle remains in the original site in the form of an isolated
disclination, avoiding a global structural transformation to move a topological
charge in crystal. The distorted lattice in the trajectory of a self-propelled
active particle is observed to restore the hexagonal configuration. Few
neutral quadrupoles may be excited in the trajectory of a moving particle to
release the slight local residue stress.  Notably, simulations show that when
an active particle switches from the random to the ballistic motion, the
originally swelled surrounding defect cluster shrinks in the direction of the
motion, leading to two types of wing-like defect motifs. We further observe the
random oscillation of a dressed active particle between these two defect motifs
regardless of the speed of the motion. The revealed uniform physical picture
of dressed active particle in this work has implications for the engineering of
defects in materials design, and it also provides the basis for further
investigation of the statistical behaviors of active particles.

\section{Model}

We construct the initial state of spherical crystal composed of point particles from a regular icosahedron
with 12 vertices and 20 triangles. In each triangle, we first introduce $n-1$
particles on each bond to equally divide the bond into $n$ segments. The
original triangle is divided by connecting any two particles at the same
height relative to their opposite side.  Extra particles are placed at the
intersect of these connecting lines. The total number of particles in the
triangulated icosahedron is $N=10(n^2-1)+12$. For a spherical crystal of $N$
particles and area $A_0$, the lattice spacing $a=\sqrt{2A_0/(\sqrt{3}N)}$.  By
mapping the vertices to a sphere whose center coincides with the center of the
icosahedron, we obtain a spherical crystal with 12 evenly distributed 5-fold
disclinations. Such a spherical crystal has a minimum number of defects allowed
by the topological constraint, and provides an ideal environment to study the
interplay of active particles and isolated defects. Note that in terms of the
Caspar$-$Klug construction for hexagonal and triangular lattices on sphere, the
lattice of our constructed spherical crystal has the coordinates of $(p, q)$
with $q=0$.~\cite{caspar1962physical} $p$ and $q$ are the numbers of steps
between successive pentagons on a spherical crystal. We will show later that
the physical picture of dressed active particle does not rely on the specific
value of $p$ and $q$.

We work in the regime of overdamped dynamics. In the hydrodynamics of small
size particles where the Reynolds number is sufficiently small, the inertial
effect can be ignored.  The motion of the particles conforms to the overdamped
Langevin equation which reads:
~\cite{szabo2006phase}
\begin{eqnarray}
    \eta \dot{\vec{r}}_i = 
    \vec{P}_{T} [ \vec{r}_i(t), f \hat{u}_i(t) + \sum_{j} \vec{F}_{ij} + A
    \vec{\xi}_i(t) ], \label{Langevin}
\end{eqnarray}
where ${\vec{r}}_i$ is the position of the particle labeled $i$
($i=1,2,3...N$). The projection operator $\vec{P}_{T}[\vec{r}_i(t), \vec{a} ] =
\vec{a} - (\hat{r}_i(t) \cdot \vec{a}) \hat{r}_i(t) $. $\eta \dot{\vec{r}}_i$
is the viscous force on the particle $i$. The three terms in the second
expression in the square bracket in Eq.(\ref{Langevin}) represent three
contributions to the forces on the particles. $f \hat{u}_i(t)$ is the
self-propulsion force of magnitude $f$ and temporally varying direction
$\hat{u}_i(t)$. About the orientation $\hat{u}_i(t)$ of the active force, we consider the
following cases: (1) the orientation of $\hat{u}_i(t)$ is generated from the
uniform distribution in the interval of $[0, 2\pi]$; (2) ballistic motion,
where $\hat{u}_i(t)$ is a constant; (3) the orientation of $\hat{u}_i(t)$
changes by angle $\theta$ in each time step, which conforms to the Gaussian
distribution with mean zero and variance $\Delta \theta^2$. For simplicity, here we
do not include the time-correlation in the evolution of $\hat{u}_i(t)$. The
time-correlation of noise is important in understanding glassy
dynamics.~\cite{szamel2015glassy}  $\vec{F}_{ij}=-\nabla V_{ij}$ is the Coulomb
force on the particle $i$ exerted by the particle $j$.  $V_{ij} = \beta
\frac{1}{|\vec{r}_i - \vec{r}_j|}$.  The last term $A \vec{\xi}_i(t)$ models
the random force on the particle $i$.  $A$ is the amplitude of the force, and
$\vec{\xi}_i(t)$ is a delta-correlated Gaussian noise. $\langle \vec{\xi}_i(t)
\rangle = 0$. $\langle \xi_i,_{\alpha}(t) \xi_j,_{\beta}(t)
\rangle = \delta_{ij}\delta_{\alpha\beta}$, where $\alpha$ and $\beta$ denote the
components of the vector $\vec{\xi}_i(t)$ in the associated tangent plane at $\vec{r}_i$.
$\vec{\xi}_i$ is in the tangent plane at ${\vec{r}}_i$ on sphere.  Note that
the noise term $\xi(t)dt$ in Eq.(\ref{Langevin}) is interpreted in the frame of
stochastic calculus.~\cite{numericalsde, ihle2014towards} Specifically, the
integration is over the standard Wiener process $\{ W(t), t \geq 0 \}$, where
$\Delta W=W(t+\Delta t)-W(t)$ constitutes independent increments of the random
variable $W(t)$. Note that the electrostatic interaction between two charged
particles whose size is much smaller than their separation dominates over their
hydrodynamic interaction~\cite{kim2005}. We therefore do not consider the hydrodynamic
interaction between particles in the model. We measure length in the unit of the lattice
spacing $a$, energy in the unit of $\epsilon_0=\beta/a$, time in the unit of
$\tau = \eta a^3/\beta$, and force in the unit of $\epsilon_0/a$.

From Eq.(\ref{Langevin}), we can construct the trajectory of the particles
from
\begin{eqnarray}
  \vec{r}_i(t+\Delta t) = \vec{r}_i(t) +  \dot{\vec{r}}_i \Delta t . \label{rt}
\end{eqnarray}
In numerically solving the Langevin equation, we choose the time step $\Delta t
= 10^{-3}\tau$. For convenience in simulations, the magnitude $A$ of the noise
is expressed in terms of $\Gamma a$, where $\Gamma$ is a fraction of unity.
$\eta \frac{\Gamma a}{\Delta t} = A$. In terms of the units for energy
($\epsilon_0$) and time ($\tau$), $A = \Gamma \epsilon_0 \tau /(a \Delta t) $.
From Eqs.(\ref{Langevin}) and (\ref{rt}), we obtain the dimensionless
discretized Langevin equation \begin{eqnarray}
\frac{\tilde{\vec{r}}_i(\tilde{t}+\Delta \tilde{t}) -
\tilde{\vec{r}}_i(\tilde{t})}{\Delta \tilde{t}} = \vec{P}_{T}
[\tilde{\vec{r}}_i(\tilde{t}), \frac{\tilde{f}}{\Delta \tilde{t}}
\hat{u}_i(\tilde{t}) + \sum_{j} \tilde{\vec{F}}_{ij} \nonumber \\  +
\frac{\Gamma}{\Delta \tilde{t}} \vec{\xi}_i(\tilde{t}) ],
\label{Langevin_dimensionless} \end{eqnarray}
where all the quantities are dimensionless. $\tilde{t}=t/\tau$.
 $\tilde{\vec{r}} = \vec{r}/a$. $\tilde{\vec{F}}= \vec{F}a/\epsilon_0$.
The activity of the particles is
controlled by $\tilde{f}a$, which is a fraction of the
lattice spacing $a$ during a time step $\Delta t$. $\tilde{f} = f \Delta t/(\eta a)$. It
is important to point out that the strength of noise and active force is
characterized by $\Gamma a$ and $\tilde{f}a$, respectively, which are fractions of the
lattice spacing $a$. The resulting factor of $\sqrt{\Delta t}$ appearing in the
integration of the standard Wiener process is absorbed in these two
dimensionless quantities.~\cite{numericalsde} In this way, the strength of both
noise and active force is well controlled in simulations.  The theoretical
model in this work may be realized experimentally in curved colloidal crystals
formed at the spherical interfaces of water and oil;~\cite{bausch2003grain}
active forces may be introduced by making use of the coupling of magnetic
colloids and controllable external magnetic field.~\cite{bausch2003grain}

\section{RESULTS AND DISCUSSION}

\begin{figure}[h]
\centering
\includegraphics[width=2.5in, bb= 400 800 2200 2000]{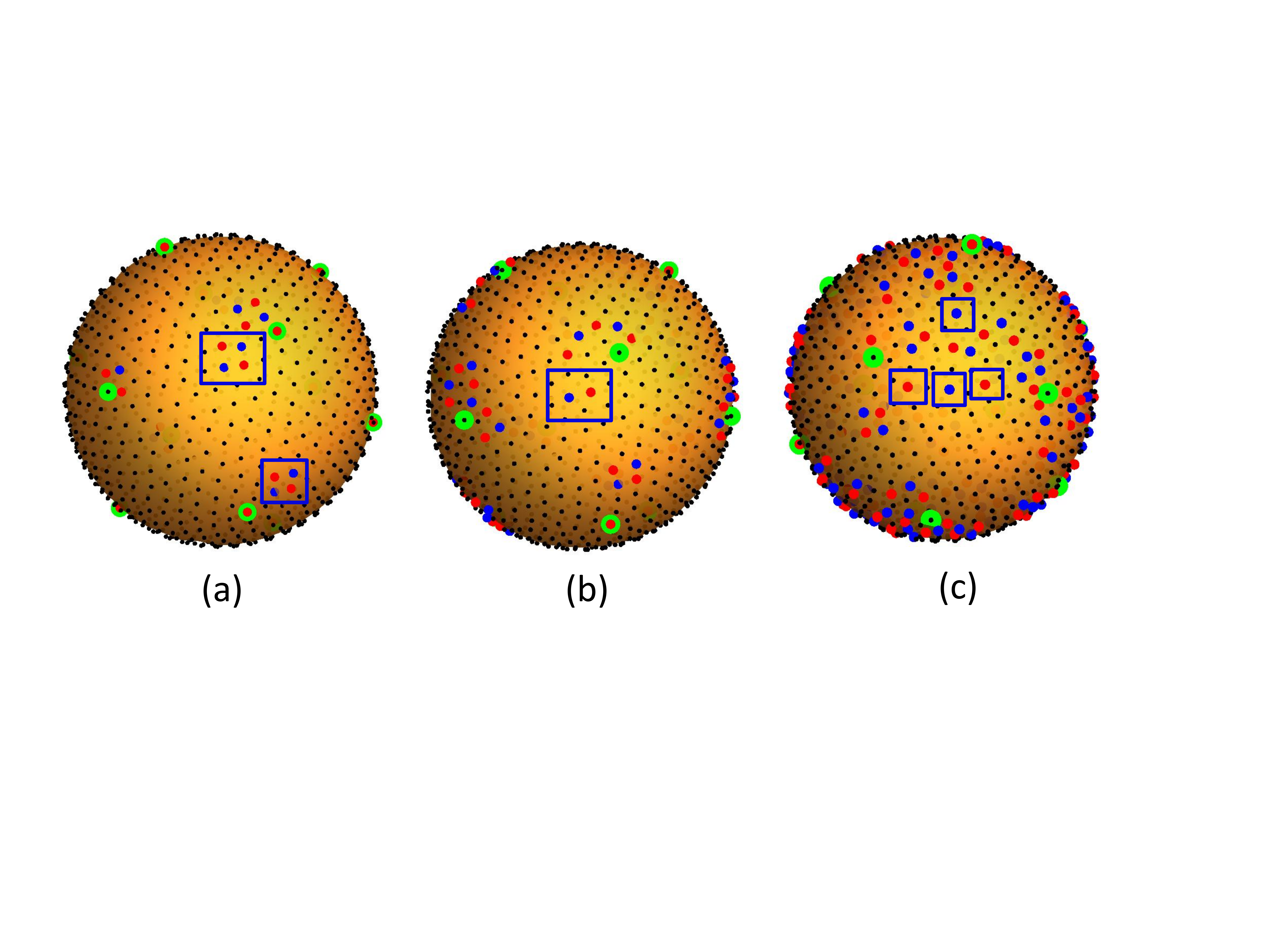}
\caption{Particles configurations at enhanced noise levels from $\Gamma =
    0.04$ (a), $\Gamma = 0.05$ (b), to $\Gamma = 0.07$ (c) without imposing any
active force. Quadrupoles [in the box in (a)], isolated dislocations [in the box in
(b)] and isolated disclinations [in the box in (c)] appear in sequence with the
increase of the noise strength $\Gamma$. The positions of the initial 12
5-fold disclinations are indicated by the large green dots. The red and blue
dots represent 5- and 7-fold disclinations. $N=1002$. 
} \label{melting}
\end{figure}

We first study the effect of the Gaussian noise in
Eq.(\ref{Langevin_dimensionless}) whose strength is characterized by the
parameter $\Gamma$. The expected melting of the spherical crystal at high level
of noise provides a qualitative criterion to check the reliability of our
simulations.  Figure~\ref{melting} shows the typical snapshots of the spherical
crystal with the increase of the noise strength $\Gamma$. Simulations capture
the splitting of quadrupoles into isolated dislocations [see
Fig.~\ref{melting}(b)] and their further fission into isolated 5- and 7-fold
disclinations [from Fig.~\ref{melting}(c)]. Therefore, the simulated spherical
crystal system experiences the intermediate hexatic phase characterized by the
proliferation of dislocations in the noise driven crystal-to-liquid phase
transition. This hexatic phase has been predicted by the  KTHNY theory for the
melting of infinitely large two-dimensional planar
crystals.~\cite{kosterlitz1973ordering, halperin1978theory,
nelson1979dislocation, young1979melting, strandburg1988two} Note that MD
simulations of two-dimensional melting on sphere with a repulsive $r^{-12}$
potential also return results that are consistent with the KTHNY
theory.~\cite{perez1998simulations} In our system, the parameter $\Gamma$ that
controls the step size in the random motion of the particles plays a similar
role of temperature in the KTHNY theory.  The appearance sequence of
quadrupoles, dislocations, and disclinations with the increase of $\Gamma$ is
also observed in smaller spherical crystals. In the spherical crystal of 252
particles, isolated dislocations and disclinations appear at
$\Gamma_{c_1}=0.04$ and $\Gamma_{c_2}=0.05$, respectively. $\Gamma_{c_1}$ is
the critical melting condition from crystal to hexatic phase, and
$\Gamma_{c_2}$ is that from hexatic to liquid phase. Both these critical values
for $\Gamma$ are smaller than those in the larger system of 1002 particles.

The size-dependence of the critical melting conditions is related to the
compactness of the spherical crystal. By confining more mutually repulsive
particles in a compact surface without a boundary like a sphere or a torus, the
system becomes stiffer.  Specifically, for $N$ electrically charged particles
with the $1/r$-Coulomb potential confined on sphere, the energy of the system
increases with $N$ in the form of $E(N) \propto N^2/2 - 0.5510
N^{3/2}$.~\cite{erber1991equilibrium} It leads to an enhanced Young's modulus
of the elastic medium composed of the equilibrium hexagonal lattice. The
binding energy of a dislocation pair and a disclination pair is proportional to
the Young’s modulus.~\cite{chaikin2000principles} Consequently, the binding
energy of the defect clusters like quadrupoles and dislocations is elevated
with the increase of the number of particles. We therefore require a larger
$\Gamma$ to activate the proliferation of isolated dislocations and
disclinations to realize the respective phase transitions.

\subsection{Random active force}

Now we tune the noise $\Gamma$ to a low level below the melting point in order
to highlight the dynamic features of an active particle.  We introduce an
active particle in the spherical crystal by imposing a random active force
$\tilde{f}$ in Eq.(\ref{Langevin_dimensionless}), and study its dynamics and
interplay with the pre-existent disclinations.  With the increase of the
magnitude of the active force, we observe that the active particle finally
escapes from the initially trapped state and gains mobility when $\tilde{f}$
exceeds some critical value $\tilde{f}_{\rm critical}$. The value for
$\tilde{f}_{\rm critical}$ is unaffected by the noise strength $\Gamma$. For a
spherical crystal of $1002$ particles, $\tilde{f}_{\rm critical} \approx 0.03$
for $\Gamma$ ranging from $0.001$ to $0.04$.  For an arbitrarily chosen trapped active particle at
low $\Gamma$, whether it is a disclination or not, increasing the noise
strength only induces a localized slight distortion of the crystal lattice near
the trapped active particle; the maximum amount of the movement of the active
particle does not exceed even one lattice spacing over up to a million time
steps. Therefore, it is the active force instead of the noise that provides the
drive force to propel a particle.

\begin{figure}[h]
\centering
\includegraphics[width=4.5in, bb= 70 20 1000 600]{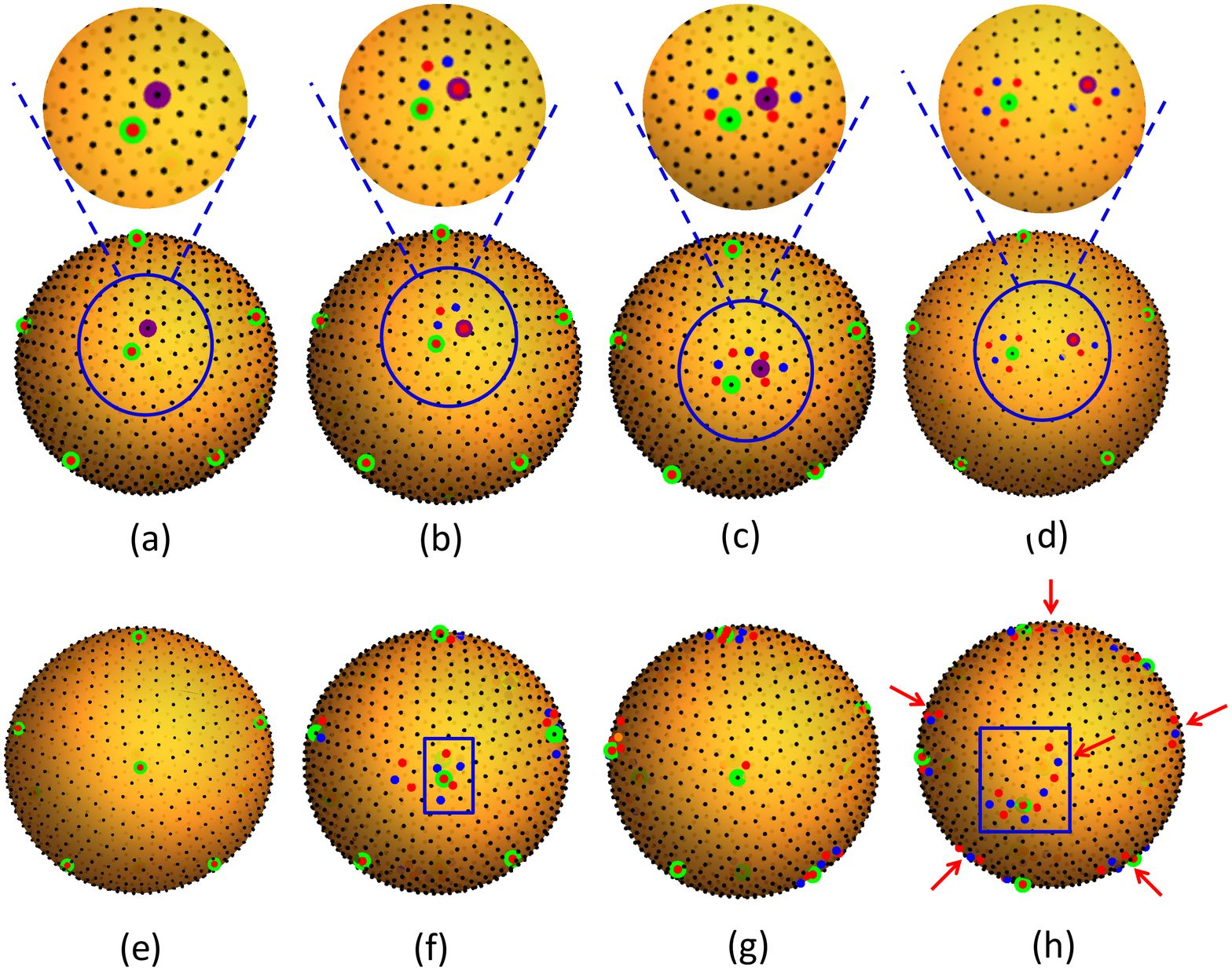}
\caption{Typical particles configurations to show the dynamic behaviors of
    active particles represented by large purple dots in (a)-(d) and
    large green dots in (e)-(h), respectively. The red and blue
dots represent 5- and 7-fold disclinations. $\tilde{f}  =0.05$.
    $\Gamma=0.001$ (a-d) and $0.002$ (e-h). $N=1002$. $t/ \Delta t
    =1100,2100,4100,8100$ from (a)-(d). $t/ \Delta t=22000,24000,28000,44000$ from
    (e)-(h).
    } \label{act_pt}
\end{figure}

Figures~\ref{act_pt}(a)-\ref{act_pt}(d) show the interaction of the active
particle (the large purple dot) initially near an isolated 5-fold disclination
(the large green dot) [see Fig.~\ref{act_pt}(a)]. We numerically observe that
the active particle induces topological charges when it approaches the isolated
disclination, creating a defect cluster with total topological charge $+1$ [see
Fig.~\ref{act_pt}(b)]. This phenomenon can be understood in the following
way.  According to the elasticity theory of topological defects, disclinations
of the same sign repel and unlike signs attract. A pair of positive and
negative disclinations constitute a dislocation which is analogous to an
electric dipole. An isolated disclination can induce the formation of
dislocations around it. The total topological charge in this process must be
invariant as a topological constraint.~\cite{chaikin2000principles} On the
other hand, it costs energy to create defects in crystal. Around an isolated
disclination, when the reduction of the energy due to the attraction of the
disclination and dislocation(s) exceeds the energy required to create
dislocation(s), the isolated disclination becomes a cluster of defects. The
transformation of an isolated disclination into a linear defect structure
called scar has been experimentally observed in spherical crystal with the
increase of the sphere radius.~\cite{bowick2006crystalline} A scar is a defect
string of alternating 5-fold and 7-fold disclinations, but with one more 5-fold
disclination.  Here, we observe that an active particle can induce such a
transformation without any change of the sphere size.

As shown in Fig.~\ref{act_pt}(c) and (d), the random active force later pulls
the active particle away from the defect cluster. When the active particle
moves in the crystalline zone among the isolated disclinations, the distorted
crystalline lattice behind it is healed.  While few neutral quadrupoles appear
in the smaller system of $N=252$, the trajectory of a moving particle in the
larger system of $N=1002$ is free of defects. This can be attributed to the
larger Young's modulus in larger systems, which elevates the energy barrier for
the formation of defects.~\cite{chaikin2000principles}  No isolated
dislocations in the trajectory of a moving particle are observed; more energy
is required to create a dislocation than a quadrupole according to the
elasticity theory of topological defects.~\cite{chaikin2000principles}

The restoration of the distorted crystal lattice in the trajectory behind
the motion of an active particle can be attributed to the long-range repulsive
interaction potential between particles. The extra free space created by a
moving particle is filled up by surrounding particles under the
long-range repulsive force. In order to confirm that a defect-free trajectory
does not rely on the specific form of the long-range force, we
perform further simulations using another long-range force in
the form of $F(r)\sim 1/r$. It turns out that such a force can also support
defect-free trajectories.  In contrast, in the two-dimensional Lennard-Jones
(L-J) crystal, where particles interact via the L-J potential, the emergent
stable vacancies may impede the complete healing of the trajectory of a
moving particle.~\cite{yao2014dynamics} Simulations show that when we remove
the attractive part in the L-J potential, the repulsive potential in the form
of $1/r^{12}$ decays so fast that it cannot support the initially prepared
crystalline order on sphere with the proliferation of defects.

To further exclude the possible influence of the dynamics of the active force
on the healing of the crystalline order in the trajectory of a moving particle,
we simulate cases where the reorientation of $\hat{u}_i(t)$ (the direction
of the active force) is subject to a finite noise instead of being completely
random. Specifically, $\hat{u}_i(t)$ rotates by angle $\theta$ in each time
step, which conforms to the Gaussian distribution with mean zero and variance
$\Delta \theta^2$. For both cases of $\Delta \theta=\pi/12$ and $\pi/6$, we
numerically observe the restoration of the distorted lattice in the
trajectory of the active particle; defects are accumulated around the moving
particle. It is of interest to note that the aggregation phenomenon of defects
is also observed in our recently studied size-polydispersity driven distortion
of crystal lattice.~\cite{yao2014} It is found that an impurity particle of a
wrong size in a perfect crystal induces localized defect
patterns to screen the effect of the impurity particle.

Figure~\ref{act_pt}(d) shows that the crystalline order very near the active
particle (in the region of a few lattice spacings) is disrupted. The active
particle carries a topologically neutral defect cluster when it moves around
without any contact with the pre-existent disclinations. It is essentially
through this defect cluster that an active particle interacts with the isolated
disclinations. Simulations show that an active particle also carries a neutral
defect cluster when moving in the lattice sufficiently away from any of the
isolated disclinations.  Therefore, the surrounding defect cluster dressing the
active particle is not caused by contact with a disclination. Such a companion
defect structure reflects the intrinsic local lattice distortion near the
active particle. This basic scenario of dressed active particle with
surrounding topological defects is also observed in systems where the active
particle is initially also an isolated disclination.

Moving an isolated disclination in crystalline order requires a global
transformation of the crystal lattice due to the topological property of the
disclination.~\cite{chaikin2000principles} Take an n-fold disclination in
two-dimensional hexagonal crystal for example, it can be created by removing
(for $n<6$) or adding (for $n>6$) a $|6-n|\pi/3$
wedge.~\cite{chaikin2000principles}  Therefore, unlike dislocations that can
freely glide across a crystalline medium, the motion of isolated disclinations
is usually realized through interactions with other defects or by evolving into
a scar to extend itself in space.~\cite{bausch2003grain} It is of interest to
endow an isolated disclination with activity and to observe the dynamics of the
active disclination. For its dual role as an active particle and also as a
disclination, here such a particle is named a disclinational active particle.
We impose sufficiently large active forces on the 12 isolated disclinations to
mobilize them [see Fig.~\ref{act_pt}(e)]. We observe the proliferation of
topological defects around the active particle and the ultimate split of the
defect cluster into a scar [the 5-7-5 configuration in the left of the blue box
in Fig.~\ref{act_pt}(f)] and a dressed active particle [the large green dot in
the blue box in Fig.~\ref{act_pt}(f)]. This fractionalization event clearly
shows that the topological feature and the activity of a particle is separable.
This phenomenon is also demonstrated in Fig.~\ref{act_pt}(g), where the
originally active disclination evolves into an isolated disclination (the red
dot in the center) and an active particle with zero topological charge (the
large green dot).

The resulting free-standing topologically charged scar [see
Fig.~\ref{act_pt}(f)] from the split of the defect cluster inherits the
topological charge from the original disclinational active particle. Both
theoretical and experimental studies on the ground state of spherical crystals
show that when $R/a$ exceeds about 5, an isolated disclination becomes a scar
to lower the energy.~\cite{bowick2002crystalline, bausch2003grain,
bowick2006crystalline} For the system in Fig.~\ref{act_pt} with $N=1002$,
$R/a=8.3$, which exceeds the critical value for the appearance of scars.
Isolated disclinations are therefore metastable.  Once touched by a dressed
active particle, a disclination becomes a scar as shown in Fig.~\ref{act_pt}(d)
and \ref{act_pt}(f). This observation shows that active particles can
facilitate the system to conquer the energy barrier to reach a new low-energy
state. In this process, the source of the required energy is from the energy
input through the active particle.  Once formed, a scar is observed to be
anchored in the spherical crystal.  The arrows in Fig.~\ref{act_pt}(h) indicate
the locations of the scars or topologically charged defect clusters, all of
which are close to the original sites of the isolated disclinations in
Fig.~\ref{act_pt}(e).    Therefore, the migration of the active particle,
whether it is initially an disclination or not, can only take away a
topologically neutral defect cluster.  The net topological charge remains in
its initial position. The origin of this phenomenon can be traced down to the
topological property of the disclination that requires a global transformation
of the crystal lattice to change its position.

The conversion of regularly distributed isolated disclinations into scars of
distinct orientations lowers the symmetry of the system. Numerical observation
of moving particles in the resulting scarred spherical crystal indicates that
the basic physical picture of dressed active particle does not rely on the
specific position and orientation of the pre-existent defects.  It can be
attributed to the fact that these defects influence the state of an active
particle only when they are as close as few lattice spacings.  The defect
pattern around the active particle is unaffected by pre-existent defects a few
lattice spacings away.  Consequently, the scenario of dressed active particle
is also expected in the less symmetric lattices with nonzero $q$ or $p \neq q$
in comparison with those of $q=0$; these less symmetric lattices essentially
change the relative orientations and positions of the disclinations.

\subsection{Ballistic active force}

In the preceding discussions, we have established that an active particle,
whether it is a disclination or not, always carries a collection of defects in
its random motion. These defects essentially reflect the localized lattice
distortion near the active particle. One may wonder if such a scenario of a
randomly wandering dressed active particle is also true in its ballistic motion
especially when it reaches a high speed. In crystalline materials,
fast moving defects can use their kinetic energy to create new defect
structures.~\cite{mott1948report} A defect at high speed
is also able to overcome obstacles such as precipitated particles or other
defects lying across its path.~\cite{cottrell1965dislocations} However, it is a
challenge to accelerate a disclination or dislocation whose speed is limited by
the sound speed in the medium in a way similar to a relativistic
particle.~\cite{cottrell1965dislocations} Fast speed of defects may be achieved
under large stress. Nevertheless, the maximum magnitude of the imposed stress
is limited by the yield stress of the material.

\begin{figure}[h]
\centering
\includegraphics[width=4.5in, bb= 50 200  1000 550]{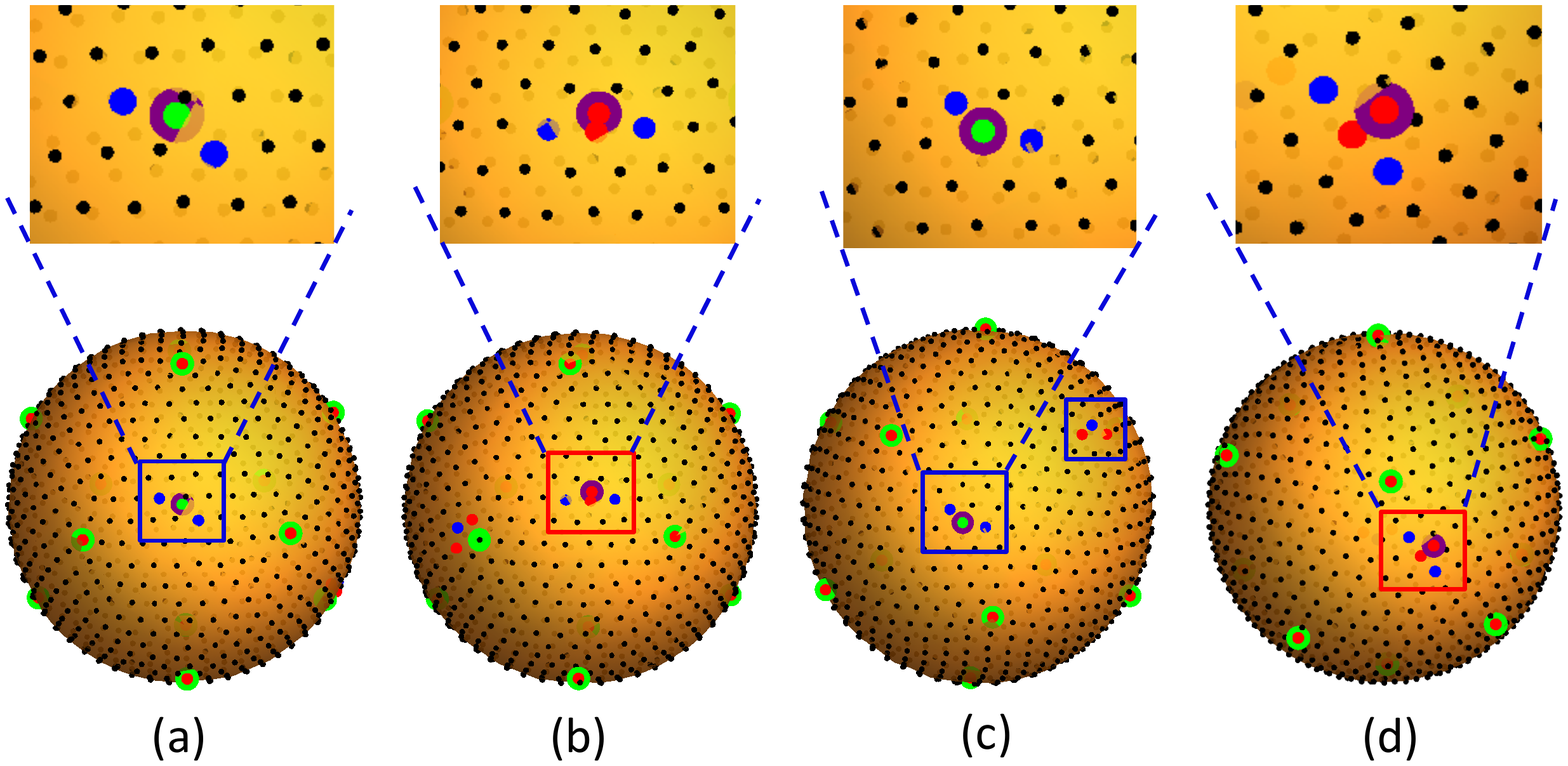}
\caption{Typical defect patterns around a ballistic active particle (the large
    purple dots). An active particle in ballistic motion, either a
    disclination (c, d) or not (a, b), oscillates between two types of defect
    patterns as shown in the boxes. The red and blue dots represent
5- and 7-fold disclinations. $\Gamma=0.002$. $\tilde{f} =0.1$ (a,b) and $0.2$
(c,d).  $t/ \Delta t=95000$ (a), $145000$ (b), $55000$ (c), and $65000$ (d). $N=1002$.
} \label{ballistic}
\end{figure}

\begin{figure}[t]
\centering
\includegraphics[width=1.3in, bb=100 20 300 250]{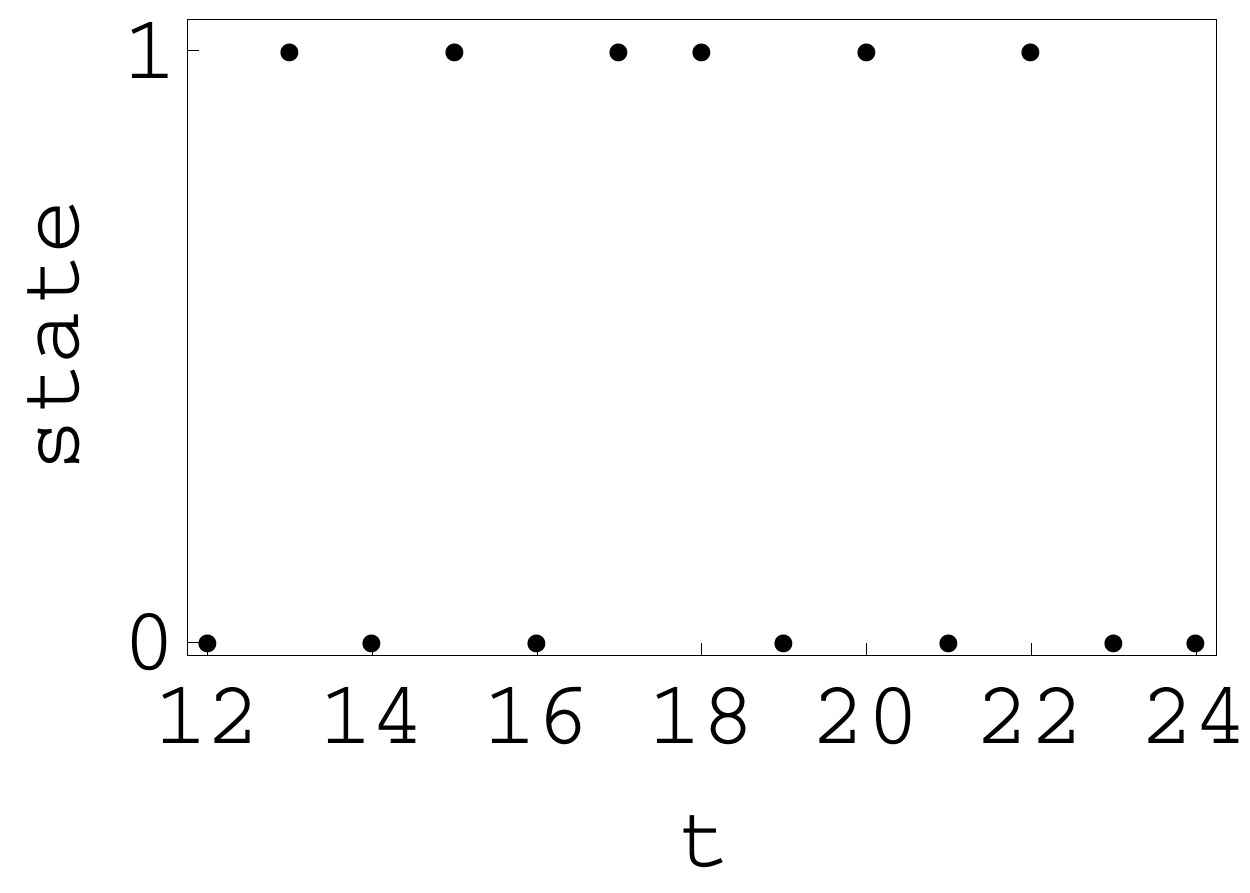}
\caption{The defect pattern surrounding an active particle in ballistic motion
oscillates between two states: state $0$ and state $1$ in the boxes in
Figs.~\ref{ballistic}(a) and \ref{ballistic}(b), respectively.  $\Gamma=0.002$.
$\tilde{f} =0.1$.  $t$ is measured in the unit of $5000\Delta t$. $N=1002$. 
} \label{two_state}
\end{figure}

\begin{figure}[t]
\centering
\includegraphics[width=3.5in, , bb=0 100 1000 500]{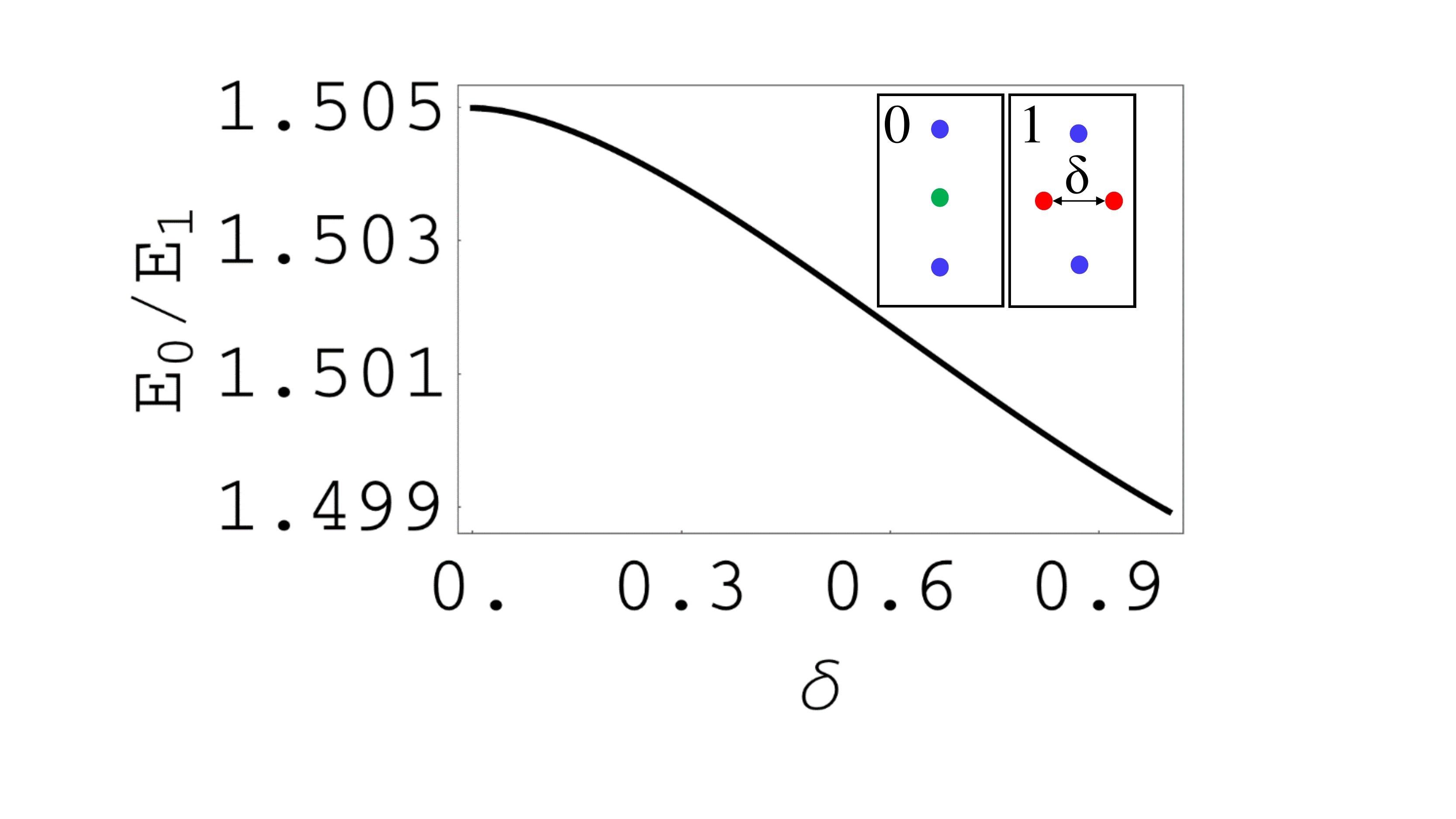}
\caption{The ratio of the energies of the two defect patterns around a
ballistic active particle. $E_0$ and $E_1$ are the energies of the state 0 and
the state 1, respectively, as shown in the inset figures (blue dot: 7-fold
disclination; green dot: 4-fold disclination; red dot: 5-fold disclination). In
these two defect configurations, disclinations of different types are separated
by a lattice spacing. } \label{two_state_energy}
\end{figure}

Our model system provides a convenient playground to inspect the idea of
employing a ballistic active particle to accelerate defects. We regulate the
motion of an active particle to be along a uniform direction at constant speed
by letting $\hat{u}_i(t)$ be a constant vector and $\tilde{f}$ a constant in
Eq.(\ref{Langevin_dimensionless}). The selected particle subject to such a
constant active force is in ballistic motion; the very low level of the noise
does not change the ballistic nature of the motion.  The speed of the active
particle can be controlled by the value for $\tilde{f}$ in
Eq.(\ref{Langevin_dimensionless}).  We first consider the case where the active
particle is not a disclination and investigate the particle speed ranging
from $\tilde{f}=0.05$ to as high as $\tilde{f}=0.5$. The previously discussed
various defect patterns formed around a randomly moving active particle are
observed to uniformly converge to either of the two well-defined topologically
neutral defect structures shown in Fig.~\ref{ballistic}(a) and
~\ref{ballistic}(b), where the active particle is represented by the large
purple dot in the box.  For convenience, in the following discussions, the
defect patterns around the active particle in Fig.~\ref{ballistic}(a) and
~\ref{ballistic}(b) are named as state 0 and state 1, respectively. In the
state 0, the disclinations are organized in the form of 7-4-7, {\it i.e.,}  a
string of 7-, 4-, and 7-fold disclinations. The state 1 is in the form of
7-(5-5)-7, where the two 5-fold disclinations can be very close and are
perpendicular to the line of the two 7-fold disclinations. The active particle
moves perpendicular to these linear defects.  Notably,the originally swelled
defect cluster dressing a randomly moving active particle shrinks in the
direction of the motion, leading to the wing-like structures.

Simulations show that in the ballistic motion of the active particle the
associated defect pattern oscillates between the two states: state $0$ (7-4-7)
in Fig.~\ref{ballistic}(a) and state $1$ [7-(5-5)-7] in
Fig.~\ref{ballistic}(b). To quantify the oscillation between these two states,
we plot Fig.~\ref{two_state} to track the dynamics of the defect pattern around
the active particle in time. For the case in Fig.~\ref{two_state} where
$\Gamma=0.002$ and $\tilde{f} =0.1$, and also for the cases of fast ($\tilde{f}
=0.5$) and slow ($\tilde{f} =0.05$) active particles, the defect pattern does
not prefer either of these two states.  Analysis of simulation data suggests
that the transition between these two states is random.

These wing-like defect motifs may become unstable with the increase of $R/a$,
where $R$ is the radius of sphere and $a$ is the lattice spacing. $R/a
\rightarrow \infty$ in the continuum limit.  It has been established
theoretically and experimentally that an isolated disclination becomes a
linear scar when $R/a$ exceeds about 5.~\cite{bowick2002crystalline,
bausch2003grain, bowick2006crystalline} The topological charge of the defect
in this transformation is preserved. For the topologically neutral wing-like
defect, it is speculated based on the case of disclination that it may become a
neutral linear defect composed of connecting dislocations. Such a defect called
pleat has been experimentally observed in curved crystals confined on
capillary bridges.~\cite{irvine2010pleats}

Here we estimate the energies of the state 0 and the state 1 of the defect pattern
around the ballistic active particle from the elasticity theory of topological
defects.~\cite{bowick2002crystalline, bausch2003grain, bowick2006crystalline} The interaction energy
between two disclinations of topological charges $q_i$ and $q_j$ in the
spherical crystal is~\cite{bowick2002crystalline} 
\begin{eqnarray} \chi(\beta) \propto q_i q_j
\left(1+\int_0^{\frac{1-\cos\beta}{2}} dz \frac{\ln z}{1-z}\right), \label{chi}
\end{eqnarray}
where $\beta$ is the angular distance between the two disclinations. The
integral in Eq.(\ref{chi}) can be expressed in terms of the polylogarithm
function $Li_n(z)=\sum_{k=1}^{\infty} z^k/k^n$: 
\begin{eqnarray} \int_0^{\frac{1-\cos\beta}{2}} dz \frac{\ln z}{1-z} =
    -\frac{\pi^2}{6}+ Li_2\left(\frac{1+\cos\beta}{2}\right), 
\end{eqnarray}
For small separation $\beta$, 
\begin{eqnarray} 
\chi(\beta) \propto q_i q_j
\left(1-\frac{1}{4}\beta^2 \ln(\frac{4e}{\beta^2}) + {\cal O}(\beta^3)   \right),
\end{eqnarray}
where $e$ is the Euler's number.  
Figure~\ref{two_state_energy} shows the ratio of $E_0$ to $E_1$ vs
$\delta$. $E_0$ and $E_1$ are the energies of the state 0 and the state 1,
respectively. $\delta$ is the separation of the two 5-fold disclinations in the
state 1 as shown in the second inset in Fig.~\ref{two_state_energy},
measured in the unit of the lattice spacing.  We see from
Fig.~\ref{two_state_energy} that the value for $E_0/E_1$ is about $3/2$, and it
is almost independent of $\delta$. Despite the appreciable energy difference of
these two states, their quasi-equal appearance frequency suggests that the
energy input to maintain a constant speed of the active particle is much larger
than their energy difference.

For a disclinational active particle in its ballistic motion, we also observe
that it leaves a net topological charge of $+1$ in its original site in the
form of a $5-7-5$ scar [see the defect in the right smaller box in
Fig.~\ref{ballistic}(c)] just like for a randomly moving disclinational
active particle. The defect pattern around the active particle in its ballistic
motion also oscillates between the state 0 [where the defects are arranged in
the form of 7-4-7 as in the large purple dot in the large box in
Fig.~\ref{ballistic}(c)] and state 1 (7-(5-5)-7) as shown in
Fig.~\ref{ballistic}(c) and ~\ref{ballistic}(d), respectively. We vary the
speed of the active particle by adjusting the value for $\tilde{f}$ from $0.01$
to $0.5$. The basic physical picture of dressed active particle and the
oscillation between the two states remains over such a broad
spectrum for the particle speed.

It is of interest to note that periodic oscillation of defects has been
experimentally observed in spherical active nematics, arising from the
collective motion of all the active microtubules composing the liquid crystal
vesicle.~\cite{keber2014topology} The conformations of the defects are
preserved in this process. In contrast, in our spherical crystal system the
defect structures are excited by an arbitrarily chosen individual particle.
And we observe featured phenomena that are absent in the collective motion in spherical
active nematics, including the healing of the defects in the trajectory of the
active particle, the scenario of dressed active particle, and the random
two-state oscillation of the surrounding defect pattern.

\

In our system of electrically charged particles confined on sphere, the
emergent defect structures are induced by the motion of
a self-propelled active particle in the crystalline order.  Similar phenomena
also occur in other physical and biological systems, where the motion of an
object in an ordered medium can excite emergent structures therein. The
specific form of the resulting structures reflects the nature of the ordered
medium. For example, a sufficiently fast moving cylinder in fluid can excite
the formation of vortices.~\cite{landau1987fluid} In crystalline materials,
fast moving defects can create new defect structures.~\cite{mott1948report} A
recent study on the collective migration of deformable biological cells shows
that individual eukaryotic cells caged in the hexagonal arrangement of cells
can deform themselves and exhibit a wiggling motion to escape from the cells
cluster. In this process, cells compete for the emergent structure of voids
which are formed in the deformation of cells.~\cite{lober2015collisions}

\section{CONCLUSION}

In summary, simulations show that a moving active particle in spherical crystal
is surrounded by localized topological defects, becoming a dressed active
particle. As a consequence of the long-range repulsion between particles, the
trajectory of the active particle is free of defects. We further observe the
random oscillation of a ballistic active particle between two defect states.
The nonequilibrium behaviors of spherical crystal excited by a moving active
particle involve dynamics of all the particles. We demonstrate that focusing
on the structure of topological defects significantly reduces the degree of
freedom of the system, and the physical picture of dressed active particle
emerges. This work opens the promising possibility of introducing activity to
efficiently engineer crystallographic defects for desired materials properties.

\section*{Acknowledgement}

This work was supported by the SJTU startup fund under Grant No. WF220441904 and the award of the Chinese
Thousand Talents Program for Distinguished Young Scholars under Grant No. 16Z127060004.

\providecommand*{\mcitethebibliography}{\thebibliography}
\csname @ifundefined\endcsname{endmcitethebibliography}
{\let\endmcitethebibliography\endthebibliography}{}


\begin{mcitethebibliography}{58}
\providecommand*{\natexlab}[1]{#1}
\providecommand*{\mciteSetBstSublistMode}[1]{}
\providecommand*{\mciteSetBstMaxWidthForm}[2]{}
\providecommand*{\mciteBstWouldAddEndPuncttrue}
  {\def\EndOfBibitem{\unskip.}}
\providecommand*{\mciteBstWouldAddEndPunctfalse}
  {\let\EndOfBibitem\relax}
\providecommand*{\mciteSetBstMidEndSepPunct}[3]{}
\providecommand*{\mciteSetBstSublistLabelBeginEnd}[3]{}
\providecommand*{\EndOfBibitem}{}
\mciteSetBstSublistMode{f}
\mciteSetBstMaxWidthForm{subitem}
{(\emph{\alph{mcitesubitemcount}})}
\mciteSetBstSublistLabelBeginEnd{\mcitemaxwidthsubitemform\space}
{\relax}{\relax}

\bibitem[Vicsek \emph{et~al.}(1995)Vicsek, Czir{\'o}k, Ben-Jacob, Cohen, and
  Shochet]{vicsek1995novel}
T.~Vicsek, A.~Czir{\'o}k, E.~Ben-Jacob, I.~Cohen and O.~Shochet, \emph{Phys.
  Rev. Lett.}, 1995, \textbf{75}, 1226\relax
\mciteBstWouldAddEndPuncttrue
\mciteSetBstMidEndSepPunct{\mcitedefaultmidpunct}
{\mcitedefaultendpunct}{\mcitedefaultseppunct}\relax
\EndOfBibitem
\bibitem[Zwanzig(2001)]{zwanzig2001nonequilibrium}
R.~Zwanzig, \emph{Nonequilibrium Statistical Mechanics}, Oxford University
  Press, USA, 2001\relax
\mciteBstWouldAddEndPuncttrue
\mciteSetBstMidEndSepPunct{\mcitedefaultmidpunct}
{\mcitedefaultendpunct}{\mcitedefaultseppunct}\relax
\EndOfBibitem
\bibitem[Marchetti \emph{et~al.}(2013)Marchetti, Joanny, Ramaswamy, Liverpool,
  Prost, Rao, and Simha]{marchetti2013hydrodynamics}
M.~Marchetti, J.~Joanny, S.~Ramaswamy, T.~Liverpool, J.~Prost, M.~Rao and R.~A.
  Simha, \emph{Rev. Mod. Phys.}, 2013, \textbf{85}, 1143\relax
\mciteBstWouldAddEndPuncttrue
\mciteSetBstMidEndSepPunct{\mcitedefaultmidpunct}
{\mcitedefaultendpunct}{\mcitedefaultseppunct}\relax
\EndOfBibitem
\bibitem[Ni \emph{et~al.}(2013)Ni, Stuart, and Dijkstra]{ni2013pushing}
R.~Ni, M.~A.~C. Stuart and M.~Dijkstra, \emph{Nat. Commun.}, 2013, \textbf{4},
  2704\relax
\mciteBstWouldAddEndPuncttrue
\mciteSetBstMidEndSepPunct{\mcitedefaultmidpunct}
{\mcitedefaultendpunct}{\mcitedefaultseppunct}\relax
\EndOfBibitem
\bibitem[Parrish and Hamner(1997)]{parrish1997animal}
J.~K. Parrish and W.~M. Hamner, \emph{Animal Groups in Three Dimensions: How
  Species Aggregate}, Cambridge University Press, 1997\relax
\mciteBstWouldAddEndPuncttrue
\mciteSetBstMidEndSepPunct{\mcitedefaultmidpunct}
{\mcitedefaultendpunct}{\mcitedefaultseppunct}\relax
\EndOfBibitem
\bibitem[Toner and Tu(1998)]{toner1998flocks}
J.~Toner and Y.~Tu, \emph{Phys. Rev. E}, 1998, \textbf{58}, 4828\relax
\mciteBstWouldAddEndPuncttrue
\mciteSetBstMidEndSepPunct{\mcitedefaultmidpunct}
{\mcitedefaultendpunct}{\mcitedefaultseppunct}\relax
\EndOfBibitem
\bibitem[Z{\"o}ttl and Stark(2016)]{zottl2016emergent}
A.~Z{\"o}ttl and H.~Stark, \emph{arXiv preprint arXiv:1601.06643}, 2016\relax
\mciteBstWouldAddEndPuncttrue
\mciteSetBstMidEndSepPunct{\mcitedefaultmidpunct}
{\mcitedefaultendpunct}{\mcitedefaultseppunct}\relax
\EndOfBibitem
\bibitem[Zhang \emph{et~al.}(2010)Zhang, Be’er, Florin, and
  Swinney]{zhang2010collective}
H.-P. Zhang, A.~Be’er, E.-L. Florin and H.~L. Swinney, \emph{Proc. Natl.
  Acad. Sci. U.S.A.}, 2010, \textbf{107}, 13626--13630\relax
\mciteBstWouldAddEndPuncttrue
\mciteSetBstMidEndSepPunct{\mcitedefaultmidpunct}
{\mcitedefaultendpunct}{\mcitedefaultseppunct}\relax
\EndOfBibitem
\bibitem[Rappel \emph{et~al.}(1999)Rappel, Nicol, Sarkissian, Levine, and
  Loomis]{rappel1999self}
W.-J. Rappel, A.~Nicol, A.~Sarkissian, H.~Levine and W.~F. Loomis, \emph{Phys.
  Rev. Lett.}, 1999, \textbf{83}, 1247\relax
\mciteBstWouldAddEndPuncttrue
\mciteSetBstMidEndSepPunct{\mcitedefaultmidpunct}
{\mcitedefaultendpunct}{\mcitedefaultseppunct}\relax
\EndOfBibitem
\bibitem[Trepat \emph{et~al.}(2009)Trepat, Wasserman, Angelini, Millet, Weitz,
  Butler, and Fredberg]{trepat2009physical}
X.~Trepat, M.~R. Wasserman, T.~E. Angelini, E.~Millet, D.~A. Weitz, J.~P.
  Butler and J.~J. Fredberg, \emph{Nat. Phys.}, 2009, \textbf{5},
  426--430\relax
\mciteBstWouldAddEndPuncttrue
\mciteSetBstMidEndSepPunct{\mcitedefaultmidpunct}
{\mcitedefaultendpunct}{\mcitedefaultseppunct}\relax
\EndOfBibitem
\bibitem[Wang and Wolynes(2011)]{wang2011spontaneous}
S.~Wang and P.~G. Wolynes, \emph{Proc. Natl. Acad. Sci. U.S.A.}, 2011,
  \textbf{108}, 15184--15189\relax
\mciteBstWouldAddEndPuncttrue
\mciteSetBstMidEndSepPunct{\mcitedefaultmidpunct}
{\mcitedefaultendpunct}{\mcitedefaultseppunct}\relax
\EndOfBibitem
\bibitem[Zhou \emph{et~al.}(2014)Zhou, Sokolov, Lavrentovich, and
  Aranson]{zhou2014living}
S.~Zhou, A.~Sokolov, O.~D. Lavrentovich and I.~S. Aranson, \emph{Proc. Natl.
  Acad. Sci. U.S.A.}, 2014, \textbf{111}, 1265--1270\relax
\mciteBstWouldAddEndPuncttrue
\mciteSetBstMidEndSepPunct{\mcitedefaultmidpunct}
{\mcitedefaultendpunct}{\mcitedefaultseppunct}\relax
\EndOfBibitem
\bibitem[Aranson and Tsimring(2006)]{aranson2006patterns}
I.~S. Aranson and L.~S. Tsimring, \emph{Rev. Mod. Phys.}, 2006, \textbf{78},
  641\relax
\mciteBstWouldAddEndPuncttrue
\mciteSetBstMidEndSepPunct{\mcitedefaultmidpunct}
{\mcitedefaultendpunct}{\mcitedefaultseppunct}\relax
\EndOfBibitem
\bibitem[Narayan \emph{et~al.}(2007)Narayan, Ramaswamy, and
  Menon]{narayan2007long}
V.~Narayan, S.~Ramaswamy and N.~Menon, \emph{Science}, 2007, \textbf{317},
  105--108\relax
\mciteBstWouldAddEndPuncttrue
\mciteSetBstMidEndSepPunct{\mcitedefaultmidpunct}
{\mcitedefaultendpunct}{\mcitedefaultseppunct}\relax
\EndOfBibitem
\bibitem[Mognetti \emph{et~al.}(2013)Mognetti, {\v{S}}ari{\'c},
  Angioletti-Uberti, Cacciuto, Valeriani, and Frenkel]{mognetti2013living}
B.~M. Mognetti, A.~{\v{S}}ari{\'c}, S.~Angioletti-Uberti, A.~Cacciuto,
  C.~Valeriani and D.~Frenkel, \emph{Phys. Rev. Lett.}, 2013, \textbf{111},
  245702\relax
\mciteBstWouldAddEndPuncttrue
\mciteSetBstMidEndSepPunct{\mcitedefaultmidpunct}
{\mcitedefaultendpunct}{\mcitedefaultseppunct}\relax
\EndOfBibitem
\bibitem[Paxton \emph{et~al.}(2004)Paxton, Kistler, Olmeda, Sen, St.~Angelo,
  Cao, Mallouk, Lammert, and Crespi]{paxton2004catalytic}
W.~F. Paxton, K.~C. Kistler, C.~C. Olmeda, A.~Sen, S.~K. St.~Angelo, Y.~Cao,
  T.~E. Mallouk, P.~E. Lammert and V.~H. Crespi, \emph{Journal of the American
  Chemical Society}, 2004, \textbf{126}, 13424--13431\relax
\mciteBstWouldAddEndPuncttrue
\mciteSetBstMidEndSepPunct{\mcitedefaultmidpunct}
{\mcitedefaultendpunct}{\mcitedefaultseppunct}\relax
\EndOfBibitem
\bibitem[Henkes \emph{et~al.}(2011)Henkes, Fily, and
  Marchetti]{henkes2011active}
S.~Henkes, Y.~Fily and M.~C. Marchetti, \emph{Phys. Rev. E}, 2011, \textbf{84},
  040301\relax
\mciteBstWouldAddEndPuncttrue
\mciteSetBstMidEndSepPunct{\mcitedefaultmidpunct}
{\mcitedefaultendpunct}{\mcitedefaultseppunct}\relax
\EndOfBibitem
\bibitem[Nguyen \emph{et~al.}(2014)Nguyen, Klotsa, Engel, and
  Glotzer]{nguyen2014emergent}
N.~H. Nguyen, D.~Klotsa, M.~Engel and S.~C. Glotzer, \emph{Phys. Rev. Lett.},
  2014, \textbf{112}, 075701\relax
\mciteBstWouldAddEndPuncttrue
\mciteSetBstMidEndSepPunct{\mcitedefaultmidpunct}
{\mcitedefaultendpunct}{\mcitedefaultseppunct}\relax
\EndOfBibitem
\bibitem[Spellings \emph{et~al.}(2015)Spellings, Engel, Klotsa, Sabrina, Drews,
  Nguyen, Bishop, and Glotzer]{spellings2015shape}
M.~Spellings, M.~Engel, D.~Klotsa, S.~Sabrina, A.~M. Drews, N.~H. Nguyen, K.~J.
  Bishop and S.~C. Glotzer, \emph{Proc. Natl. Acad. Sci. U.S.A.}, 2015,
  \textbf{112}, E4642--E4650\relax
\mciteBstWouldAddEndPuncttrue
\mciteSetBstMidEndSepPunct{\mcitedefaultmidpunct}
{\mcitedefaultendpunct}{\mcitedefaultseppunct}\relax
\EndOfBibitem
\bibitem[Schweitzer and Farmer(2007)]{schweitzer2007brownian}
F.~Schweitzer and J.~Farmer, \emph{Brownian Agents and Active Particles},
  Springer, 2007\relax
\mciteBstWouldAddEndPuncttrue
\mciteSetBstMidEndSepPunct{\mcitedefaultmidpunct}
{\mcitedefaultendpunct}{\mcitedefaultseppunct}\relax
\EndOfBibitem
\bibitem[Vicsek and Zafeiris(2012)]{vicsek2012collective}
T.~Vicsek and A.~Zafeiris, \emph{Phys. Rep.}, 2012, \textbf{517}, 71--140\relax
\mciteBstWouldAddEndPuncttrue
\mciteSetBstMidEndSepPunct{\mcitedefaultmidpunct}
{\mcitedefaultendpunct}{\mcitedefaultseppunct}\relax
\EndOfBibitem
\bibitem[Tierno \emph{et~al.}(2008)Tierno, Golestanian, Pagonabarraga, and
  Sagu{\'e}s]{tierno2008controlled}
P.~Tierno, R.~Golestanian, I.~Pagonabarraga and F.~Sagu{\'e}s, \emph{Phys. Rev.
  Lett.}, 2008, \textbf{101}, 218304\relax
\mciteBstWouldAddEndPuncttrue
\mciteSetBstMidEndSepPunct{\mcitedefaultmidpunct}
{\mcitedefaultendpunct}{\mcitedefaultseppunct}\relax
\EndOfBibitem
\bibitem[Palacci \emph{et~al.}(2013)Palacci, Sacanna, Steinberg, Pine, and
  Chaikin]{palacci2013living}
J.~Palacci, S.~Sacanna, A.~P. Steinberg, D.~J. Pine and P.~M. Chaikin,
  \emph{Science}, 2013, \textbf{339}, 936--940\relax
\mciteBstWouldAddEndPuncttrue
\mciteSetBstMidEndSepPunct{\mcitedefaultmidpunct}
{\mcitedefaultendpunct}{\mcitedefaultseppunct}\relax
\EndOfBibitem
\bibitem[Palacci \emph{et~al.}(2013)Palacci, Sacanna, Vatchinsky, Chaikin, and
  Pine]{palacci2013photoactivated}
J.~Palacci, S.~Sacanna, A.~Vatchinsky, P.~M. Chaikin and D.~J. Pine, \emph{J.
  Am. Chem. Soc.}, 2013, \textbf{135}, 15978--15981\relax
\mciteBstWouldAddEndPuncttrue
\mciteSetBstMidEndSepPunct{\mcitedefaultmidpunct}
{\mcitedefaultendpunct}{\mcitedefaultseppunct}\relax
\EndOfBibitem
\bibitem[Zhang \emph{et~al.}(2014)Zhang, Walker, Grzybowski, and Olvera de~la
  Cruz]{zhang2014accelerated}
R.~Zhang, D.~A. Walker, B.~A. Grzybowski and M.~Olvera de~la Cruz, \emph{Angew.
  Chem. Int. Ed.}, 2014, \textbf{126}, 177--181\relax
\mciteBstWouldAddEndPuncttrue
\mciteSetBstMidEndSepPunct{\mcitedefaultmidpunct}
{\mcitedefaultendpunct}{\mcitedefaultseppunct}\relax
\EndOfBibitem
\bibitem[Chaudhuri \emph{et~al.}(2011)Chaudhuri, Bhattacharya, Gowrishankar,
  Mayor, and Rao]{chaudhuri2011spatiotemporal}
A.~Chaudhuri, B.~Bhattacharya, K.~Gowrishankar, S.~Mayor and M.~Rao,
  \emph{Proc. Natl. Acad. Sci. U.S.A.}, 2011, \textbf{108}, 14825--14830\relax
\mciteBstWouldAddEndPuncttrue
\mciteSetBstMidEndSepPunct{\mcitedefaultmidpunct}
{\mcitedefaultendpunct}{\mcitedefaultseppunct}\relax
\EndOfBibitem
\bibitem[Sanchez \emph{et~al.}(2012)Sanchez, Chen, DeCamp, Heymann, and
  Dogic]{sanchez2012spontaneous}
T.~Sanchez, D.~T. Chen, S.~J. DeCamp, M.~Heymann and Z.~Dogic, \emph{Nature},
  2012, \textbf{491}, 431--434\relax
\mciteBstWouldAddEndPuncttrue
\mciteSetBstMidEndSepPunct{\mcitedefaultmidpunct}
{\mcitedefaultendpunct}{\mcitedefaultseppunct}\relax
\EndOfBibitem
\bibitem[Shi and Ma(2013)]{shi2013topological}
X.-Q. Shi and Y.-Q. Ma, \emph{Nat. Commun.}, 2013, \textbf{4}, 3013\relax
\mciteBstWouldAddEndPuncttrue
\mciteSetBstMidEndSepPunct{\mcitedefaultmidpunct}
{\mcitedefaultendpunct}{\mcitedefaultseppunct}\relax
\EndOfBibitem
\bibitem[Giomi \emph{et~al.}(2013)Giomi, Bowick, Ma, and
  Marchetti]{giomi2013defect}
L.~Giomi, M.~J. Bowick, X.~Ma and M.~C. Marchetti, \emph{Phys. Rev. Lett.},
  2013, \textbf{110}, 228101\relax
\mciteBstWouldAddEndPuncttrue
\mciteSetBstMidEndSepPunct{\mcitedefaultmidpunct}
{\mcitedefaultendpunct}{\mcitedefaultseppunct}\relax
\EndOfBibitem
\bibitem[Sknepnek and Henkes(2015)]{sknepnek2015active}
R.~Sknepnek and S.~Henkes, \emph{Phys. Rev. E}, 2015, \textbf{91}, 022306\relax
\mciteBstWouldAddEndPuncttrue
\mciteSetBstMidEndSepPunct{\mcitedefaultmidpunct}
{\mcitedefaultendpunct}{\mcitedefaultseppunct}\relax
\EndOfBibitem
\bibitem[Schaller and Bausch(2013)]{schaller2013topological}
V.~Schaller and A.~R. Bausch, \emph{Proc. Natl. Acad. Sci. U.S.A.}, 2013,
  \textbf{110}, 4488--4493\relax
\mciteBstWouldAddEndPuncttrue
\mciteSetBstMidEndSepPunct{\mcitedefaultmidpunct}
{\mcitedefaultendpunct}{\mcitedefaultseppunct}\relax
\EndOfBibitem
\bibitem[Penrose(1965)]{penrose1965dermatoglyphic}
L.~Penrose, \emph{Nature}, 1965, \textbf{205}, 544--546\relax
\mciteBstWouldAddEndPuncttrue
\mciteSetBstMidEndSepPunct{\mcitedefaultmidpunct}
{\mcitedefaultendpunct}{\mcitedefaultseppunct}\relax
\EndOfBibitem
\bibitem[Keber \emph{et~al.}(2014)Keber, Loiseau, Sanchez, DeCamp, Giomi,
  Bowick, Marchetti, Dogic, and Bausch]{keber2014topology}
F.~C. Keber, E.~Loiseau, T.~Sanchez, S.~J. DeCamp, L.~Giomi, M.~J. Bowick,
  M.~C. Marchetti, Z.~Dogic and A.~R. Bausch, \emph{Science}, 2014,
  \textbf{345}, 1135--1139\relax
\mciteBstWouldAddEndPuncttrue
\mciteSetBstMidEndSepPunct{\mcitedefaultmidpunct}
{\mcitedefaultendpunct}{\mcitedefaultseppunct}\relax
\EndOfBibitem
\bibitem[Bowick \emph{et~al.}(2002)Bowick, Cacciuto, Nelson, and
  Travesset]{bowick2002crystalline}
M.~Bowick, A.~Cacciuto, D.~R. Nelson and A.~Travesset, \emph{Phys. Rev. Lett.},
  2002, \textbf{89}, 185502\relax
\mciteBstWouldAddEndPuncttrue
\mciteSetBstMidEndSepPunct{\mcitedefaultmidpunct}
{\mcitedefaultendpunct}{\mcitedefaultseppunct}\relax
\EndOfBibitem
\bibitem[Bausch \emph{et~al.}(2003)Bausch, Bowick, Cacciuto, Dinsmore, Hsu,
  Nelson, Nikolaides, Travesset, and Weitz]{bausch2003grain}
A.~Bausch, M.~Bowick, A.~Cacciuto, A.~Dinsmore, M.~Hsu, D.~Nelson,
  M.~Nikolaides, A.~Travesset and D.~Weitz, \emph{Science}, 2003, \textbf{299},
  1716--1718\relax
\mciteBstWouldAddEndPuncttrue
\mciteSetBstMidEndSepPunct{\mcitedefaultmidpunct}
{\mcitedefaultendpunct}{\mcitedefaultseppunct}\relax
\EndOfBibitem
\bibitem[Bowick \emph{et~al.}(2006)Bowick, Cacciuto, Nelson, and
  Travesset]{bowick2006crystalline}
M.~J. Bowick, A.~Cacciuto, D.~R. Nelson and A.~Travesset, \emph{Phys. Rev. B},
  2006, \textbf{73}, 024115\relax
\mciteBstWouldAddEndPuncttrue
\mciteSetBstMidEndSepPunct{\mcitedefaultmidpunct}
{\mcitedefaultendpunct}{\mcitedefaultseppunct}\relax
\EndOfBibitem
\bibitem[Struik(1988)]{struik88a}
D.~Struik, \emph{Lectures on Classical Differential Geometry}, Dover
  Publications, 2nd edn, 1988\relax
\mciteBstWouldAddEndPuncttrue
\mciteSetBstMidEndSepPunct{\mcitedefaultmidpunct}
{\mcitedefaultendpunct}{\mcitedefaultseppunct}\relax
\EndOfBibitem
\bibitem[Chaikin and Lubensky(2000)]{chaikin2000principles}
P.~M. Chaikin and T.~C. Lubensky, \emph{Principles of Condensed Matter
  Physics}, Cambridge Univ Press, 2000, vol.~1\relax
\mciteBstWouldAddEndPuncttrue
\mciteSetBstMidEndSepPunct{\mcitedefaultmidpunct}
{\mcitedefaultendpunct}{\mcitedefaultseppunct}\relax
\EndOfBibitem
\bibitem[Kosterlitz and Thouless(1973)]{kosterlitz1973ordering}
J.~M. Kosterlitz and D.~J. Thouless, \emph{J. Phys. C: Solid State Phys.},
  1973, \textbf{6}, 1181\relax
\mciteBstWouldAddEndPuncttrue
\mciteSetBstMidEndSepPunct{\mcitedefaultmidpunct}
{\mcitedefaultendpunct}{\mcitedefaultseppunct}\relax
\EndOfBibitem
\bibitem[Halperin and Nelson(1978)]{halperin1978theory}
B.~Halperin and D.~R. Nelson, \emph{Phys. Rev. Lett.}, 1978, \textbf{41},
  121\relax
\mciteBstWouldAddEndPuncttrue
\mciteSetBstMidEndSepPunct{\mcitedefaultmidpunct}
{\mcitedefaultendpunct}{\mcitedefaultseppunct}\relax
\EndOfBibitem
\bibitem[Nelson and Halperin(1979)]{nelson1979dislocation}
D.~R. Nelson and B.~Halperin, \emph{Phys. Rev. B}, 1979, \textbf{19},
  2457\relax
\mciteBstWouldAddEndPuncttrue
\mciteSetBstMidEndSepPunct{\mcitedefaultmidpunct}
{\mcitedefaultendpunct}{\mcitedefaultseppunct}\relax
\EndOfBibitem
\bibitem[Young(1979)]{young1979melting}
A.~Young, \emph{Phys. Rev. B}, 1979, \textbf{19}, 1855\relax
\mciteBstWouldAddEndPuncttrue
\mciteSetBstMidEndSepPunct{\mcitedefaultmidpunct}
{\mcitedefaultendpunct}{\mcitedefaultseppunct}\relax
\EndOfBibitem
\bibitem[Strandburg(1988)]{strandburg1988two}
K.~J. Strandburg, \emph{Rev. Mod. Phys.}, 1988, \textbf{60}, 161\relax
\mciteBstWouldAddEndPuncttrue
\mciteSetBstMidEndSepPunct{\mcitedefaultmidpunct}
{\mcitedefaultendpunct}{\mcitedefaultseppunct}\relax
\EndOfBibitem
\bibitem[Caspar and Klug(1962)]{caspar1962physical}
D.~L. Caspar and A.~Klug, Cold Spring Harbor Symposia on Quantitative Biology,
  1962, pp. 1--24\relax
\mciteBstWouldAddEndPuncttrue
\mciteSetBstMidEndSepPunct{\mcitedefaultmidpunct}
{\mcitedefaultendpunct}{\mcitedefaultseppunct}\relax
\EndOfBibitem
\bibitem[Szabo \emph{et~al.}(2006)Szabo, Sz{\"o}ll{\"o}si, G{\"o}nci,
  Jur{\'a}nyi, Selmeczi, and Vicsek]{szabo2006phase}
B.~Szabo, G.~Sz{\"o}ll{\"o}si, B.~G{\"o}nci, Z.~Jur{\'a}nyi, D.~Selmeczi and
  T.~Vicsek, \emph{Phys. Rev. E}, 2006, \textbf{74}, 061908\relax
\mciteBstWouldAddEndPuncttrue
\mciteSetBstMidEndSepPunct{\mcitedefaultmidpunct}
{\mcitedefaultendpunct}{\mcitedefaultseppunct}\relax
\EndOfBibitem
\bibitem[Szamel \emph{et~al.}(2015)Szamel, Flenner, and
  Berthier]{szamel2015glassy}
G.~Szamel, E.~Flenner and L.~Berthier, \emph{Phys. Rev. E}, 2015, \textbf{91},
  062304\relax
\mciteBstWouldAddEndPuncttrue
\mciteSetBstMidEndSepPunct{\mcitedefaultmidpunct}
{\mcitedefaultendpunct}{\mcitedefaultseppunct}\relax
\EndOfBibitem
\bibitem[Kloeden and Platen(1995)]{numericalsde}
P.~Kloeden and E.~Platen, \emph{Numerical Solution of Stochastic Differential
  Equations}, Springer, 1995\relax
\mciteBstWouldAddEndPuncttrue
\mciteSetBstMidEndSepPunct{\mcitedefaultmidpunct}
{\mcitedefaultendpunct}{\mcitedefaultseppunct}\relax
\EndOfBibitem
\bibitem[Ihle(2014)]{ihle2014towards}
T.~Ihle, \emph{Eur. Phys. J. Special Topics}, 2014, \textbf{7},
  1293--1314\relax
\mciteBstWouldAddEndPuncttrue
\mciteSetBstMidEndSepPunct{\mcitedefaultmidpunct}
{\mcitedefaultendpunct}{\mcitedefaultseppunct}\relax
\EndOfBibitem
\bibitem[Kim and Karrila(2005)]{kim2005}
S.~Kim and S.~Karrila, \emph{Microhydrodynamics: Principles and Selected
  Applications}, Dover, New York, 2005\relax
\mciteBstWouldAddEndPuncttrue
\mciteSetBstMidEndSepPunct{\mcitedefaultmidpunct}
{\mcitedefaultendpunct}{\mcitedefaultseppunct}\relax
\EndOfBibitem
\bibitem[P{\'e}rez-Garrido and Moore(1998)]{perez1998simulations}
A.~P{\'e}rez-Garrido and M.~Moore, \emph{Physical Review B}, 1998, \textbf{58},
  9677\relax
\mciteBstWouldAddEndPuncttrue
\mciteSetBstMidEndSepPunct{\mcitedefaultmidpunct}
{\mcitedefaultendpunct}{\mcitedefaultseppunct}\relax
\EndOfBibitem
\bibitem[Erber and Hockney(1991)]{erber1991equilibrium}
T.~Erber and G.~Hockney, \emph{Journal of Physics A: Mathematical and General},
  1991, \textbf{24}, L1369\relax
\mciteBstWouldAddEndPuncttrue
\mciteSetBstMidEndSepPunct{\mcitedefaultmidpunct}
{\mcitedefaultendpunct}{\mcitedefaultseppunct}\relax
\EndOfBibitem
\bibitem[Yao and Olvera de~la Cruz(2014)]{yao2014dynamics}
Z.~Yao and M.~Olvera de~la Cruz, \emph{Phys. Rev. E}, 2014, \textbf{90},
  062318\relax
\mciteBstWouldAddEndPuncttrue
\mciteSetBstMidEndSepPunct{\mcitedefaultmidpunct}
{\mcitedefaultendpunct}{\mcitedefaultseppunct}\relax
\EndOfBibitem
\bibitem[Yao and Olvera de~la Cruz(2014)]{yao2014}
Z.~Yao and M.~Olvera de~la Cruz, \emph{Proc. Natl. Acad. Sci.}, 2014,
  \textbf{111}, 5094\relax
\mciteBstWouldAddEndPuncttrue
\mciteSetBstMidEndSepPunct{\mcitedefaultmidpunct}
{\mcitedefaultendpunct}{\mcitedefaultseppunct}\relax
\EndOfBibitem
\bibitem[Mott and Nabarro(1948)]{mott1948report}
N.~Mott and F.~Nabarro, \emph{Report on Strength of Solids}, Physical Society,
  London, 1948, pp. 1--19\relax
\mciteBstWouldAddEndPuncttrue
\mciteSetBstMidEndSepPunct{\mcitedefaultmidpunct}
{\mcitedefaultendpunct}{\mcitedefaultseppunct}\relax
\EndOfBibitem
\bibitem[Cottrell(1965)]{cottrell1965dislocations}
A.~H. Cottrell, \emph{Dislocations and Plastic Flow in Crystals}, Clarendon
  Press, 1965\relax
\mciteBstWouldAddEndPuncttrue
\mciteSetBstMidEndSepPunct{\mcitedefaultmidpunct}
{\mcitedefaultendpunct}{\mcitedefaultseppunct}\relax
\EndOfBibitem
\bibitem[Irvine \emph{et~al.}(2010)Irvine, Vitelli, and
  Chaikin]{irvine2010pleats}
W.~T. Irvine, V.~Vitelli and P.~M. Chaikin, \emph{Nature}, 2010, \textbf{468},
  947--951\relax
\mciteBstWouldAddEndPuncttrue
\mciteSetBstMidEndSepPunct{\mcitedefaultmidpunct}
{\mcitedefaultendpunct}{\mcitedefaultseppunct}\relax
\EndOfBibitem
\bibitem[Landau and Lifshitz(1987)]{landau1987fluid}
L.~D. Landau and E.~M. Lifshitz, \emph{Fluid Mechanics}, Butterworth,
  1987\relax
\mciteBstWouldAddEndPuncttrue
\mciteSetBstMidEndSepPunct{\mcitedefaultmidpunct}
{\mcitedefaultendpunct}{\mcitedefaultseppunct}\relax
\EndOfBibitem
\bibitem[L{\"o}ber \emph{et~al.}(2015)L{\"o}ber, Ziebert, and
  Aranson]{lober2015collisions}
J.~L{\"o}ber, F.~Ziebert and I.~S. Aranson, \emph{Sci. Rep.}, 2015, \textbf{5},
  9172\relax
\mciteBstWouldAddEndPuncttrue
\mciteSetBstMidEndSepPunct{\mcitedefaultmidpunct}
{\mcitedefaultendpunct}{\mcitedefaultseppunct}\relax
\EndOfBibitem
\end{mcitethebibliography}
\end{document}